\documentclass{vgtc}                          % final (conference style)

\ifpdf%                                % if we use pdflatex
  \pdfoutput=1\relax                   % create PDFs from pdfLaTeX
  \usepackage{graphicx}                % allow us to embed graphics files
  \DeclareGraphicsExtensions{.pdf,.png,.jpg,.jpeg} % for pdflatex we expect .pdf, .png, or .jpg files
\else%                                 % else we use pure latex
  \usepackage{graphicx}                % allow us to embed graphics files
  \DeclareGraphicsExtensions{.eps}     % for pure latex we expect eps files
\fi%

\graphicspath{{figures/}{pictures/}{images/}{./}} % where to search for the images

\usepackage{microtype}                 % use micro-typography (slightly more compact, better to read)
\PassOptionsToPackage{warn}{textcomp}  % to address font issues with \textrightarrow
\usepackage{textcomp}                  % use better special symbols
\usepackage{mathptmx}                  % use matching math font
\usepackage{times}                     % we use Times as the main font
\usepackage{cite}                      % needed to automatically sort the references
\usepackage{tabu}                      % only used for the table example
\usepackage{booktabs}                  % only used for the table example
\usepackage{graphicx}
\usepackage{booktabs}
\usepackage{multirow}
\usepackage{textcomp}
\usepackage{array}
\usepackage{cite}
\usepackage{pgfplots}
\usepackage{csquotes}
\onlineid{0}

\vgtccategory{Research}

\vgtcinsertpkg

\title{Virtual Reality Therapy for the Psychological Well-being of Palliative Care Patients in Hong Kong}

 \author{Daniel Eckhoff\thanks{e-mail: daniel.eckhoff@gmail.com} \thanks{Contributed equally}\\ %
         \parbox{1.4in}{\scriptsize \centering School of Creative Media \\ City University of Hong Kong, Hong Kong} %
 \and Royce Ng\thanks{e-mail: soloroyce@gmail.com}  \footnotemark[2]\\ 
     \parbox{1.4in}{\scriptsize \centering School of Creative Media \\ City University of Hong Kong, Hong Kong} %
 \and Alvaro Cassinelli\thanks{e-mail: cassinelli.alvaro@gmail.com}\\ %
      \parbox{1.4in}{\scriptsize \centering School of Creative Media \\ City University of Hong Kong, Hong Kong}}
\abstract{
In this paper we introduce novel Virtual Reality (VR) and Augmented Reality (AR) treatments to improve the psychological well being of patients in palliative care, based on interviews with a clinical psychologist who has successfully implemented VR assisted interventions on palliative care patients in the Hong Kong hospital system. Our VR and AR assisted interventions are adaptations of traditional palliative care therapies which simultaneously facilitate patients communication with family and friends while isolated in hospital due to physical weakness and COVID-19 related restrictions. The first system we propose is a networked, metaverse platform for palliative care patients to create customized virtual environments with therapists, family and friends which function as immersive and collaborative versions of ‘life review’ and ‘reminiscence therapy’. The second proposed system will investigate the use of Mixed Reality telepresence and haptic touch in an AR environment, which will allow palliative care patients to physically feel friends and family in a virtual space, adding to the sense of presence and immersion in that environment. %The final section of this paper will discuss the design and methodology for the pilot and field studies we will implement for use with palliative care patients. 
} 

\CCScatlist{
  \CCScatTwelve{Applied computing}{Health Informatics}{}{};
  \CCScatTwelve{Human-centered computing}{Virtual reality}{}{}
}

\begin{document}

%% The ``\maketitle'' command must be the first command after the
%% ``\begin{document}'' command. It prepares and prints the title block.

%% the only exception to this rule is the \firstsection command
\maketitle

\section{Introduction}
 In the current Hong Kong hospital system palliative care patients are often socially isolated from family and friends in their final weeks and months of life as a result of weakness and physical deterioration, a situation which has been exacerbated by strict COVID-19 social distancing measures. These circumstances often leave patients in states of psychological, emotional and spiritual distress before death, which also increases their experience of physical pain. According to a recent systematic review on the unmet supportive care needs of advanced cancer patients and their caregivers \cite{wang2020}, the most commonly reported domain was lack of psychological support, with the absence of family and friends (9.9–96.5\%) and emotional assistance (10.1–84.4\%) accounting for the largest share of unmet needs, with an additional frequently collected response being patients inability to communicate with the outside world (7.7–87.9\%).  
Since unmet needs have a detrimental effect on well-being \cite{instituteofmedicineus2008,newell1999}, an innovative intervention that is able to address all of these psychological problems while overcoming the constraints of physical isolation is of pressing demand. Palliative care plays an essential role in these circumstances, providing holistic care which aims to manage physical symptoms while addressing the psychological and emotional needs of the patient using various therapeutic methods to help them ‘prepare for death and affirm the patient as a human with a unique life.’ \cite{keall2014}. One novel and effective technique used by clinical psychologists is ‘life review’ and 'reminiscence' therapy \cite{ando2006,wang2018,warth2019}, which involves the systematic examination of various life experiences with the aim of resolving conflict and completion of life tasks in order to help the patient come to terms with death. Examples of the therapeutic outcomes of 'life review' and 'reminiscence' therapy are the writing of texts in collaboration with a therapist and the production of picture books, audio recordings or cookbooks as a 'legacy' gift for family members after the patients death. A constraint of this therapy is that it consists entirely of spoken, textual or 2D visual communication and generally does not directly involve friends and family members who were participants in the patient's life experiences. 
 
 In this respect, we propose that metaverse and VR/AR technologies offer novel and unprecedented opportunities for the fulfillment of palliative care patients unmet psychological and physical needs. We suggest that these platforms and systems can improve the psychological efficacy of 'life review' and 'reminiscence' therapy techniques at the same time as facilitating remote communication with family and friends with a greater sense of presence and immersion.
 
 This research project involves the development of a VR and an AR system designed to assist with the psychological interventions currently being implemented on palliative care patients in the Hong Kong hospital system. The first system is a virtual reality, metaverse platform for palliative care patients to communicate with friends and family using custom 3D generated environments based on emotionally significant spaces for the patient that also incorporate photos and videos from their social media accounts. These virtual social spaces will have a positive therapeutic function as embodied and networked versions of ‘life review’ and ‘reminiscence’ therapy which also allow patients to communicate and collaborate with family and friends in creating positive spaces for memory and reflection on their lives. The second proposed system will involve the integration of haptic touch into VR and AR \emph{telepresence} systems which will allow patients to not only personally communicate, but physically interact with family and friends through the use of AR holograms and haptic sensors, in which patients can meet virtual visitors in any environment or even teleport to their former home. We propose that an AR telepresence application can use haptic technologies to facilitate co-presence and enable affective communication by being able to convey a wealth of socio-emotional information and elicit many positive emotions similar to those of natural touch \cite{gallace2010}. This research project explores the unique potential for VR and AR technologies to provide palliative care patients with platforms for social interaction with the outside world which will ultimately benefit their psychological wellbeing. It will also examine dimensions for social interaction that traverse not only space through remote communication, but also time, by creating customized virtual environments that will serve as posthumous archives of patients' experiences and memories in the metaverse before and even after death. 
 
 Based on the related work and an interview with a pallitiave care expert [\ref{section:interview}] we want to address the following three research questions in our future work:
\textbf{RQ 1:} Can interactive virtual platforms for palliative care patients assist traditional life review/reminiscence therapy techniques used by clinical psychologists to improve quality of life (QOL)? 
\textbf{RQ 2}: Is there increased co-presence in a networked 3D environment presented in virtual reality for immersive communication between palliative care patients and family members compared with remote communication using 2D interfaces such as phones and laptops? 
\textbf{RQ 3:} Can haptic interaction in a networked 3D virtual environment improve the social presence and affective communication between palliative care patients with family and friends? 

To address these research questions, we are proposing and outlining the development of Virtual Reality system specifically designed for palliative care patients. Additionally, we interviewed a clinical psychologist who is already using Virtual Reality therapies for palliative care.

\section{Related Work}
A recent review conducted by Blomstrom et al. \cite{blomstrom2022} of innovative interventions used by clinical psychologists in palliative care therapy included virtual reality alongside psychotherapeutic techniques, mindfulness and psychedelics as promising new areas for research. A number of these studies have examined the use of virtual reality to overcome fear of death and dying, using near-death-experiences (NDE) which harness the full body illusions possible in VR and AR to induce out-of-body experiences which have demonstrated decreased fear-of-death outcomes for participants \cite{bourdin2017,bourdin2020}. 

There is a small but growing body of research on the the use and effectiveness of virtual reality in palliative care, with a systematic meta-analysis conducted by Mo et al. \cite{mo2022a} having examined all eight studies produced so far concluding that while the technique is feasible and acceptable, the quality of the extant studies was graded as very low. Several of the papers analyzed involve patients experiencing guided walks in nature based virtual reality environments \cite{groninger2021a}, \cite{banos2013a}, \cite{ferguson2020a}, two feature personalized experiences and virtual travel \cite{niki2019}, \cite{perna2021} and another two include music therapy and viewing of still images \cite{brungardt2021a}, \cite{johnson2020a}. From the review, two out of the eight studies reported lower pain scores amongst patients after a guided VR experience \cite{groninger2021a}, \cite{perna2021} while in six of the studies patients reported positive attitudes to using virtual reality \cite{groninger2021a}, \cite{banos2013a}, \cite{brungardt2021a}, \cite{dang2021} and four reported difficulties in operating the software and hardware \cite{banos2013a}, \cite{brungardt2021a}, \cite{ferguson2020a}, \cite{johnson2020a}. 

The interchangeably used 'life review' and 'reminiscence' therapy techniques have been increasingly utilized by clinical psychologists and studies have revealed its significant positive impact on the quality of life of palliative care patients. In the systematic review conducted by Keall et al.\cite{keall2015a} they described the diverse variety of life review therapies used by psychologist which variously include; One week short term life review that features the creation of a picture book of life events, good and bad memories accompanied with text \cite{ando2006}. Dignity therapy involving the writing of a book about the patients life which becomes a legacy gift to family \cite{chochinov2005a},\cite{chochinov2011a}. Legacy Activities in which patients work with a carer to produce legacy activities to be enjoyed by family members after death, e.g., audiotape of family stories, scrapbook, cookbook \cite{allen2008a}. 

A significant study for our research included in the systematic review was the \emph{The Meaning of Life Intervention} conducted by Mok et al. \cite{mok2012a} which involved  semi-structured  recorded interviews with patients asking the questions 'What do you think about your life?', 'How have you faced adversity in your life?' and 'What do you do to love yourself and others? What brings you joy? What do you appreciate in your life?’ The process of conducting these interviews and asking basic questions about life and mortality produced demonstrable improvements in quality of life for patients. The only study so far to incorporate virtual reality into life review therapy was conducted by Dang et al. \cite{dang2021} which developed a virtual avatar to lead a life review therapy session with palliative care patients, which bears similarity to research conducted by Tominari et al. \cite{tominari2021} which developed a virtual reminiscence therapy for dementia sufferers using 360 videos presented on touchscreen pads. In contrast to these examples, our research instead focuses on the creation of a platform for collaborative exchange between the patient, family and friends and the therapist to generate a virtual space of collective memory analogous to the legacy outcomes which result from traditional life review and reminiscence therapy interventions.    
VR/AR telepresence applications can also make an important contribution to palliative care by using haptic technologies to facilitate co-presence and enable affective communication by providing the opportunity for patients to touch their loved ones. Even relatively simple tactile actuators can be used to convey a wealth of socio-emotional information and elicit many positive emotions similar to those of natural touch \cite{gallace2010}. For example, virtual touch has been shown to reduce anxiety triggered by watching a sad video \cite{cabibihan2012} and increase shared conversation while watching a comedy clip with other users \cite{kerdvibulvech2019}. In addition to affective modulation, receiving virbrotactile signals from other users has been shown to promote co-presence and interpersonal connectedness \cite{nakanishi2014} and elicit helpful behavior and generosity in social decision-making paradigms \cite{haans2009a}.

\section{Interview with a Clinical Psychologist using Virtual Reality in Palliative Care Interventions} \label{section:interview}

Currently, there are no quantitative and qualitative data on the feasibility of using virtual reality in palliative care in Hong Kong. Therefore, we conducted a semi-structured interview with Ms. Olive Woo, one of the first clinical psychologists currently using virtual reality in her therapeutic interventions with palliative care patients in the Hong Kong public hospital system. Our questions were organized into three themes: 

\begin{enumerate}
    \item Motivations for using virtual reality with palliative care patients. 
    \item Feasibility for the use of virtual reality with palliative care patients. 
    \item The results of using virtual reality based palliative care interventions.
\end{enumerate}

\subsubsection*{Motivations for using virtual reality with palliative patients}
The wish of many terminally ill patients is to meet and say goodbye to their family and friends. However, during the COVID-19 pandemic and due to geographical restrictions, this is often not possible. For example, palliative care patients in Hong Kong currently only allows limited visitations, and Hong Kong still has strict border restrictions. Virtual and augmented reality can allow patients to meet family and friends in any virtual environment or even teleport into their own homes. Ms. Woo stated her motivations for using the technology:
\begin{displayquote}
There has been unmet needs among palliative care patients, despite the effectiveness of pharmaceutical and psychological interventions. For example, some patients wish to come back home during their last days of life, but they are not allowed to given their physical condition, some patients wish to travel to new places, but it is impossible. 
\end{displayquote}
In addition, the use of VR by Ms. Woo also responded to patients stated desires for experiences beyond their physical confinement as well as spiritually motivated desires related to their own mortality, which customized VR and AR experiences could possibly provide access to. As Ms. Woo suggests; 
\begin{displayquote}
Due to their specific needs, some contents are [not] yet available. For example, they would like to see some images related to heaven for relaxation.
\end{displayquote}
\subsubsection*{Feasibility of using virtual reality with palliative care patients}

The use of VR and AR proves advantageous in palliative care. A feasibility study by Appel and colleagues \cite{appel2020} demonstrated that older adults with reduced sensory, mobility, and/or impaired cognition could benefit from VR experiences. In the multi-site non-randomized interventional study, all sixty-six participants with mild to severe cognitive and/or physical impairments were able to complete 3 to 20 minutes of 360\textdegree VR experience. Furthermore, the results concluded that older adults had tolerated the VR experience well, with most participants exhibiting positive emotions and reflecting positive feedback, verifying VR as a practical medium to deliver an embodied experience for communication.

However, while VR and AR offer enormous potential for remote communication with greater feelings of presence and immersion, there are some side effects and dangers of the technologies which need to be considered and avoided, especially for use with the particularly vulnerable populations of palliative care patients. Common side-effects of VR and AR include cybersickness (nausea, dizziness, headache, sweating, eye strain) \cite{lawson2021a} while users, especially elderly ones, often have difficulty navigating the hardware and software \cite{banos2013a}, \cite{brungardt2021a}, \cite{ferguson2020a}, \cite{johnson2020a}. Ms. Woo used different methodological strategies to mitigate possible negative side effects which included extensive screening of patients so that unsuitable ones were excluded from the interventions to avoid unnecessary harm. 

\begin{displayquote}
Some of them wish to use VR but their physical condition does not allow. For example, tetraplegic patients or patients with cord compression may not be able to sit but lay on bed, which renders the use of VR difficult (as they can only view the upper visual field). Some of them cannot move their head, which limits the use of VR. Some of them may have motion sickness. 
\end{displayquote}

\subsubsection*{The results of using virtual reality based palliative care interventions}
As the literature review above demonstrates, the use of virtual reality in palliative care is acceptable and feasible, with some data showing its effectiveness in reducing pain while two early studies have provided evidence that VR can assist with traditional palliative care therapies such as life review and reminiscence therapy. Ms. Woo reported positive results for her VR interventions, and based on informally administered outcome measures, patients have shown;
\begin{displayquote}
Positive change in mood and physical symptoms including fatigue, pain, drowsiness, appetite.
\end{displayquote}
While Ms. Woo also expressed her positive opinion for the efficacy of virtual reality as a palliative care therapy as it provides; 
\begin{displayquote}
An easily-accessible alternative to meet their unmet needs which have yet to be addressed by traditional treatment approaches.
\end{displayquote}

\section{Future Work}
Based on the related work and the interview with the clinical psychologist, we propose a research methodology to answer our three research questions:

\subsection*{Construction of Replica Hospital Room}
In the experiments conducted using VR to induce out-of-body experiences for participants to overcome fear-of-death done by Bourdin et al. \cite{bourdin2017}, \cite{bourdin2017b}, their studies participants were healthy rather than dying patients. Following this model, we will conduct a pilot study using the systems we develop on healthy participants to rigorously test that they are safe to use before proceeding with a field study with palliative care patients. 
We will build a full-size replica of a normal hospital room for palliative patients at the City University of Hong Kong. This room will be equipped with a patient bed and a chair and couch for visitors. Since some palliative patients are often bedridden and too weak to sit, a replica room with a patient bed is crucial to evaluate the effectiveness, feasibility and user experience of our proposed systems before embarking on field studies. Therefore, we will conduct two of our pilot studies in this room on students and healthy elderly people.

\subsubsection*{Hardware and Software Development}
We will develop two software and one hardware project to answer our three proposed research questions. 

A crucial aspect for both systems is accessibility for bedridden, weak patients. For this, we will additionally develop interaction methods using eye-gaze and common VR controllers. Since some patients are too weak to even move their head and therefore unable to change their view in the virtual environment, we will develop methods to change the view without triggering motion sickness. These interaction methods will first be tested on healthy students as part of our pilot study.

For \textbf{RQ 1}, we will develop a VR Unity application for the Meta Quest 2 that will have two modes: An authoring mode that allows psychologists, caregivers, and family members to place images and videos of a palliative care patient's life in an immersive virtual environment. They will be able to choose from a variety of virtual environments, the first offers a selection of locations based on socially and culturally significant sites for the patient in Hong Kong which will provide familiar and nostalgic spaces for the patient to feel comfortable in. The second is a menu interface which allows the patient to customize their own space, using a modular selection of architecture, decor and furniture. The third provides a selection of \emph{neutral} relaxation spaces based on nature sites in Hong Kong.

In the therapy mode, the application allows the interaction with the palliative care patient to walk through the environment and see images from the patient's life. In these spaces, the patient can communicate and collaborate in real-time using avatars with their therapist, family and friends to create \emph{legacy} rooms filled with photos and videos from the patients and families social media that function as virtual and embodied versions of life review and reminiscence therapy.     

For \textbf{RQ2} and \textbf{RQ3} we will develop two Augmented Reality telepresence systems: The first system will be built for the Meta Quest 2 that allow patients and visitors to see and talk to each other via a live stream (video, audio) from 360° cameras (Ricoh Theta Z1, Ricoh Company, Ltd., Japan). Each camera transmits the live feed via a connected computer to an RTMP server in the City University of Hong Kong. The headsets receive the RTMP stream from the other end. 

\begin{figure}
    \centering
    \includegraphics[width = \linewidth]{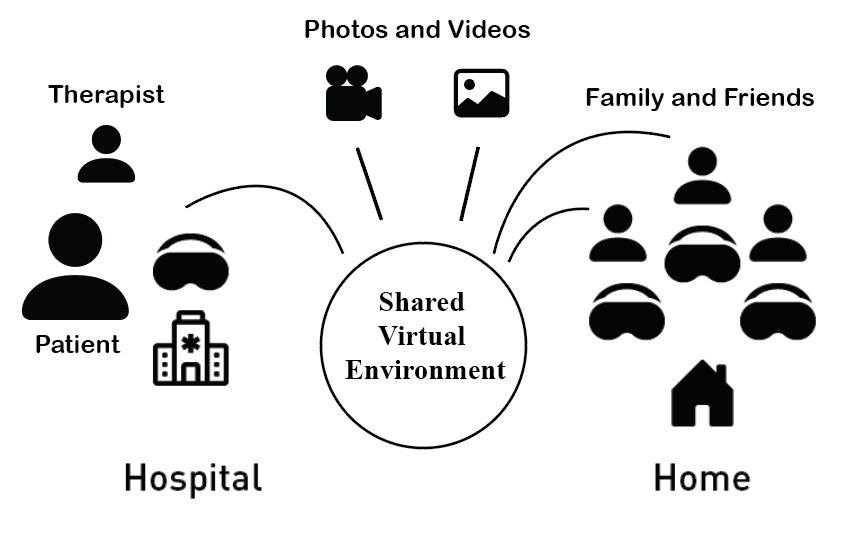}
    \centering
    \includegraphics[width = \linewidth]{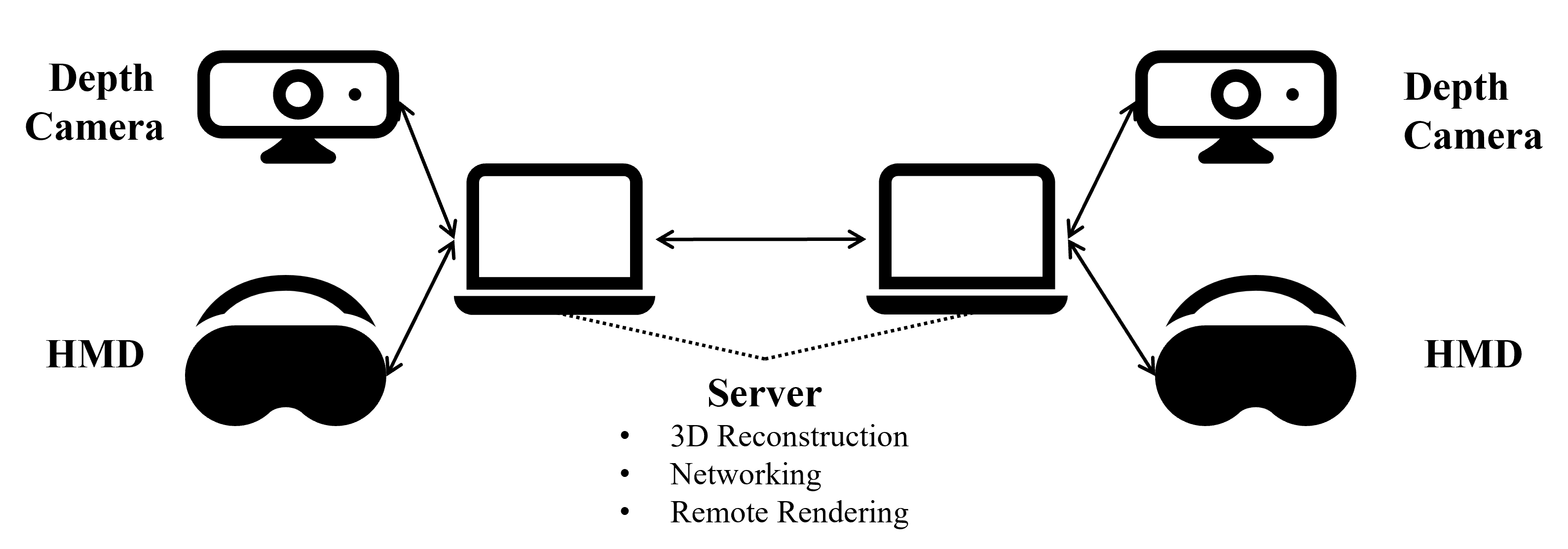}
    \caption{Top: Overview of virtual life review for palliative care system architecture. Bottom: Overview of the mixed reality telepresence system architecture.}
    \label{fig:system}
\end{figure}

The second approach is based on holoportation \cite{orts-escolano2016}. We will build a holoportation system in Unity for a Video See-Through Head-Mounted Display such as the Varjo XR-3 (Varjo Technologies Oy, Helsinki, Finland) or the Lynx R1 (Lynx Mixed Reality, Paris, France). We will use Azure Kinects (Microsoft Corporation, United States) as depth cameras, that allow high-quality body segmentation, which allows each end to see, hear as well as touch the other end. 

To implement haptic sensations when patients touch the hologram, we will develop a haptic glove that we will equip with multiple linear resonance actuators (LRA) that will generate haptic sensations on each finger and on the palm. The LRAs of the haptic glove will be controlled by an ESP32 micro controller that can communicate to the Meta Quest 2 over Bluetooth.

\subsubsection*{Pilot Study on Healthy Participants}
The pilot study consists of two parts. In the first part, we will recruit 10 healthy participants from the student population to test the reliability and user experience of our systems as well to collect feedback on our software and hardware platform. Participants will be compensated with 200 Hong Kong dollars.
After making adjustments to our platform, we will recruit 10 healthy older participants through social media ads as well as university channels to collect preliminary data for our research questions. 
\subsubsection*{Field Study - Palliative Care}
Following the pilot study and the approval of an ethical application, medical experts from
the Hong Kong Hospital Authority will collect data on patients in palliative care using our systems. Based on our pilot study, we will conduct a power analysis to determine the minimum sample size to find significant treatment effects.

\subsubsection*{Instruments}
The outcomes of \textbf{RQ1} which examines whether co-presence in a networked, virtual version of life review/reminiscence therapy can improve quality of life will be measured using the Quality-of-Life Concerns in the End-of-Life (QOLC-E) questionnaire which was adapted by Pang et al. \cite{pang2005} to record the QOL of Chinese patients with advanced-stage cancer. 

The outcomes of \textbf{RQ2} which investigate whether co-presence in a networked 3D environment presented in virtual reality for immersive communication between palliative care patients and family members compared with remote communication using 2D interfaces will be measured using a Cantonese adaptation and translation of the Presence Questionnaire (PD) developed by Slater et al. \cite{slater1995}, \cite{slater1998} to first measure the sense of personal presence experienced by the participants during the VR experience. A creation of Presence in Virtual Reality Environments (CVEs) questionnaire will then be developed in Cantonese to measure the co-presence felt by the participants during the experiment. Upon translating and creating the questionnaires, a pilot study using students will be conducted to measure its validity and reliability before its adoption for palliative care patients and their family members.

Lastly, the outcomes of \textbf{RQ3} to inspect whether haptic interaction in a networked 3D virtual environment improves the social presence and affective communication between palliative care patients with family and friends will be measured using an adaption of the questionnaire using a bipolar seven-point Likert scale from Sallnäs’ thesis \cite{sallnas2004}. The questionnaire is to be translated into Cantonese. Similar to the above questionnaires, the translated questionnaire will first be standardized using students before its adoption for the palliative care participants and their family members.

\section{Conclusion}
 Palliative patients suffer from a number of unmet needs, such as the inability to communicate with the outside world, the inability to remain at home during the last days of life, and the inability to travel with loved ones. Based on an interview with a clinical psychologist in Hong Kong who uses VR in palliative care and a literature review of the current state of virtual reality therapies for palliative care we propose that VR and AR offer easily-accessible alternative therapies to meet the needs of patients which have yet to be addressed by traditional treatment approaches. We propose three systems to improve the interventions currently in use alongside detailed methodologies for conducting a pilot and field study using these systems on healthy and palliative care patients in Hong Kong.

\section{Acknowledgements}
The authors would like to acknowledge the generous help of Ms. Olive Woo in agreeing to be interviewed and for contributing her valuable comments and expertise. We would also like to thank Ms. Vivian Yuen who assisted with the instruments section.  

\bibliographystyle{abbrv-doi}

\bibliography{library}

\begin{thebibliography}{10}

\bibitem{allen2008a}
R.~S. Allen, M.~M. Hilgeman, M.~A. Ege, J.~L. Shuster, Jr, and L.~D. Burgio.
\newblock Legacy activities as interventions approaching the end of life.
\newblock {\em Journal of palliative medicine}, 11(7):1029--1038, 2008.

\bibitem{ando2006}
M.~Ando, A.~Tsuda, and S.~Moorey.
\newblock Preliminary {{Study}} of {{Reminiscence Therapy}} on {{Depression}}
  and {{Self-Esteem}} in {{Cancer Patients}}.
\newblock {\em Psychological Reports}, 98(2):339--346, Apr. 2006. doi: {{%
10\hspace{.1pt}\discretionary{.}{%
}{.}\hspace{.4pt}2466\discretionary{/}{%
}{/}pr0\hspace{.1pt}\discretionary{.}{%
}{.}\hspace{.4pt}98\hspace{.1pt}\discretionary{.}{%
}{.}\hspace{.4pt}2\hspace{.1pt}\discretionary{.}{%
}{.}\hspace{.4pt}339\discretionary{%
}{-}{-}346}}


\bibitem{appel2020}
L.~Appel, E.~Appel, O.~Bogler, M.~Wiseman, L.~Cohen, N.~Ein, H.~Abrams, and
  J.~Campos.
\newblock Older adults with cognitive and/or physical impairments can benefit
  from immersive virtual reality experiences: A feasibility study.
\newblock {\em Frontiers in Medicine}, 6:329, 01 2020. doi: {{%
10\hspace{.1pt}\discretionary{.}{%
}{.}\hspace{.4pt}3389\discretionary{/}{%
}{/}fmed\hspace{.1pt}\discretionary{.}{%
}{.}\hspace{.4pt}2019\hspace{.1pt}\discretionary{.}{%
}{.}\hspace{.4pt}00329}}


\bibitem{banos2013a}
R.~M. Ba{\~n}os, M.~Espinoza, A.~{Garc{\'i}a-Palacios}, J.~M. Cervera,
  G.~Esquerdo, E.~Barraj{\'o}n, and C.~Botella.
\newblock A positive psychological intervention using virtual reality for
  patients with advanced cancer in a hospital setting: A pilot study to assess
  feasibility.
\newblock {\em Supportive Care in Cancer}, 21(1):263--270, Jan. 2013. doi: {{%
10\hspace{.1pt}\discretionary{.}{%
}{.}\hspace{.4pt}1007\discretionary{/}{%
}{/}s00520\discretionary{%
}{-}{-}012\discretionary{%
}{-}{-}1520\discretionary{%
}{-}{-}x}}


\bibitem{blomstrom2022}
M.~Blomstrom, A.~Burns, D.~Larriviere, and J.~K. Penberthy.
\newblock Addressing fear of death and dying: Traditional and innovative
  interventions.
\newblock {\em Mortality}, 27(1):18--37, Jan. 2022. doi: {{%
10\hspace{.1pt}\discretionary{.}{%
}{.}\hspace{.4pt}1080\discretionary{/}{%
}{/}13576275\hspace{.1pt}\discretionary{.}{%
}{.}\hspace{.4pt}2020\hspace{.1pt}\discretionary{.}{%
}{.}\hspace{.4pt}1810649}}


\bibitem{bourdin2017}
P.~Bourdin, I.~Barberia, R.~Oliva, and M.~Slater.
\newblock A {{Virtual Out-of-Body Experience Reduces Fear}} of {{Death}}.
\newblock {\em PLOS ONE}, 12(1):e0169343, Jan. 2017. doi: {{%
10\hspace{.1pt}\discretionary{.}{%
}{.}\hspace{.4pt}1371\discretionary{/}{%
}{/}journal\hspace{.1pt}\discretionary{.}{%
}{.}\hspace{.4pt}pone\hspace{.1pt}\discretionary{.}{%
}{.}\hspace{.4pt}0169343}}


\bibitem{bourdin2017b}
P.~Bourdin, I.~Barberia, R.~Oliva, and M.~Slater.
\newblock A {{Virtual Out-of-Body Experience Reduces Fear}} of {{Death}}.
\newblock {\em PLOS ONE}, 12(1):e0169343, Jan. 2017. doi: {{%
10\hspace{.1pt}\discretionary{.}{%
}{.}\hspace{.4pt}1371\discretionary{/}{%
}{/}journal\hspace{.1pt}\discretionary{.}{%
}{.}\hspace{.4pt}pone\hspace{.1pt}\discretionary{.}{%
}{.}\hspace{.4pt}0169343}}


\bibitem{bourdin2020}
P.~Bourdin, L.~Calvet, S.~Tesconi, and J.~{Arnedo-Moreno}.
\newblock Reflecting on {{Attitudes Towards Death Through}} the use of
  {{Immersive Virtual Reality Commercial Video Games}}.
\newblock In {\em Eighth {{International Conference}} on {{Technological
  Ecosystems}} for {{Enhancing Multiculturality}}}, {{TEEM}}'20, pp. 640--647.
  {Association for Computing Machinery}, {New York, NY, USA}, Oct. 2020. doi:
  {{%
10\hspace{.1pt}\discretionary{.}{%
}{.}\hspace{.4pt}1145\discretionary{/}{%
}{/}3434780\hspace{.1pt}\discretionary{.}{%
}{.}\hspace{.4pt}3436559}}


\bibitem{brungardt2021a}
A.~Brungardt, A.~Wibben, A.~F. Tompkins, P.~Shanbhag, H.~Coats, A.~B. LaGasse,
  D.~Boeldt, J.~Youngwerth, J.~S. Kutner, and H.~D. Lum.
\newblock Virtual {{Reality-Based Music Therapy}} in {{Palliative Care}}: {{A
  Pilot Implementation Trial}}.
\newblock {\em Journal of palliative medicine}, 24(5):736--742, 2021.

\bibitem{cabibihan2012}
J.-J. Cabibihan, L.~Zheng, and C.~K.~T. Cher.
\newblock Affective {{Tele-touch}}.
\newblock In D.~Hutchison, T.~Kanade, J.~Kittler, J.~M. Kleinberg, F.~Mattern,
  J.~C. Mitchell, M.~Naor, O.~Nierstrasz, C.~Pandu~Rangan, B.~Steffen,
  M.~Sudan, D.~Terzopoulos, D.~Tygar, M.~Y. Vardi, G.~Weikum, S.~S. Ge,
  O.~Khatib, J.-J. Cabibihan, R.~Simmons, and M.-A. Williams, eds., {\em Social
  {{Robotics}}}, vol. 7621, pp. 348--356. {Springer Berlin Heidelberg},
  {Berlin, Heidelberg}, 2012. doi: {{%
10\hspace{.1pt}\discretionary{.}{%
}{.}\hspace{.4pt}1007\discretionary{/}{%
}{/}978\discretionary{%
}{-}{-}3\discretionary{%
}{-}{-}642\discretionary{%
}{-}{-}34103\discretionary{%
}{-}{-}8\_35}}


\bibitem{chochinov2005a}
H.~M. Chochinov, T.~Hack, T.~Hassard, L.~J. Kristjanson, S.~McClement, and
  M.~Harlos.
\newblock Dignity {{Therapy}}: {{A Novel Psychotherapeutic Intervention}} for
  {{Patients Near}} the {{End}} of {{Life}}.
\newblock {\em Journal of Clinical Oncology}, 23(24):5520--5525, Aug. 2005.

\bibitem{chochinov2011a}
H.~M. Chochinov, L.~J. Kristjanson, W.~Breitbart, S.~McClement, T.~F. Hack,
  T.~Hassard, and M.~Harlos.
\newblock Effect of dignity therapy on distress and end-of-life experience in
  terminally ill patients: A randomised controlled trial.
\newblock {\em The Lancet Oncology}, 12(8):753--762, Aug. 2011. doi: {{%
10\hspace{.1pt}\discretionary{.}{%
}{.}\hspace{.4pt}1016\discretionary{/}{%
}{/}S1470\discretionary{%
}{-}{-}2045\discretionary{%
}{(}{(}11\discretionary{)}{%
}{)}70153\discretionary{%
}{-}{-}X}}


\bibitem{dang2021}
M.~Dang, D.~Noreika, S.~Ryu, A.~Sima, H.~Ashton, B.~Ondris, F.~Coley,
  J.~Nestler, and E.~D. Fabbro.
\newblock Feasibility of {{Delivering}} an {{Avatar-Facilitated Life Review
  Intervention}} for {{Patients}} with {{Cancer}}.
\newblock {\em Journal of Palliative Medicine}, 24(4):520--526, Apr. 2021. doi:
  {{%
10\hspace{.1pt}\discretionary{.}{%
}{.}\hspace{.4pt}1089\discretionary{/}{%
}{/}jpm\hspace{.1pt}\discretionary{.}{%
}{.}\hspace{.4pt}2020\hspace{.1pt}\discretionary{.}{%
}{.}\hspace{.4pt}0020}}


\bibitem{ferguson2020a}
C.~Ferguson, M.~Y. Shade, J.~Blaskewicz~Boron, E.~Lyden, and N.~A. Manley.
\newblock Virtual {{Reality}} for {{Therapeutic Recreation}} in {{Dementia
  Hospice Care}}: {{A Feasibility Study}}.
\newblock {\em American Journal of Hospice and Palliative
  Medicine\textregistered}, 37(10):809--815, Oct. 2020. doi: {{%
10\hspace{.1pt}\discretionary{.}{%
}{.}\hspace{.4pt}1177\discretionary{/}{%
}{/}1049909120901525}}


\bibitem{gallace2010}
A.~Gallace and C.~Spence.
\newblock The science of interpersonal touch: {{An}} overview.
\newblock {\em Neuroscience \& Biobehavioral Reviews}, 34(2):246--259, Feb.
  2010. doi: {{%
10\hspace{.1pt}\discretionary{.}{%
}{.}\hspace{.4pt}1016\discretionary{/}{%
}{/}j\hspace{.1pt}\discretionary{.}{%
}{.}\hspace{.4pt}neubiorev\hspace{.1pt}\discretionary{.}{%
}{.}\hspace{.4pt}2008\hspace{.1pt}\discretionary{.}{%
}{.}\hspace{.4pt}10\hspace{.1pt}\discretionary{.}{%
}{.}\hspace{.4pt}004}}


\bibitem{groninger2021a}
H.~Groninger, D.~Stewart, J.~M. Fisher, E.~Tefera, J.~Cowgill, and M.~Mete.
\newblock Virtual reality for pain management in advanced heart failure: {{A}}
  randomized controlled study.
\newblock {\em Palliative Medicine}, 35(10):2008--2016, Dec. 2021. doi: {{%
10\hspace{.1pt}\discretionary{.}{%
}{.}\hspace{.4pt}1177\discretionary{/}{%
}{/}02692163211041273}}


\bibitem{haans2009a}
A.~Haans and W.~A. IJsselsteijn.
\newblock The {{Virtual Midas Touch}}: {{Helping Behavior After}} a {{Mediated
  Social Touch}}.
\newblock {\em IEEE Transactions on Haptics}, 2(3):136--140, July 2009. doi:
  {{%
10\hspace{.1pt}\discretionary{.}{%
}{.}\hspace{.4pt}1109\discretionary{/}{%
}{/}TOH\hspace{.1pt}\discretionary{.}{%
}{.}\hspace{.4pt}2009\hspace{.1pt}\discretionary{.}{%
}{.}\hspace{.4pt}20}}


\bibitem{instituteofmedicineus2008}
{Institute of Medicine (US)}.
\newblock {\em Evidence-{{Based Medicine}} and the {{Changing Nature}} of
  {{Healthcare}}: 2007 {{IOM Annual Meeting Summary}}}.
\newblock The {{National Academies Collection}}: {{Reports}} Funded by
  {{National Institutes}} of {{Health}}. {National Academies Press (US)},
  {Washington (DC)}, 2008.

\bibitem{johnson2020a}
T.~Johnson, L.~Bauler, D.~Vos, A.~Hifko, P.~Garg, M.~Ahmed, and M.~Raphelson.
\newblock Virtual {{Reality Use}} for {{Symptom Management}} in {{Palliative
  Care}}: {{A Pilot Study}} to {{Assess User Perceptions}}.
\newblock {\em Journal of palliative medicine}, 23(9):1233--1238, 2020.

\bibitem{keall2014}
R.~Keall, J.~M. Clayton, and P.~Butow.
\newblock How do {{Australian}} palliative care nurses address existential and
  spiritual concerns? {{Facilitators}}, barriers and strategies.
\newblock {\em Journal of Clinical Nursing}, 23(21-22):3197--3205, 2014. doi:
  {{%
10\hspace{.1pt}\discretionary{.}{%
}{.}\hspace{.4pt}1111\discretionary{/}{%
}{/}jocn\hspace{.1pt}\discretionary{.}{%
}{.}\hspace{.4pt}12566}}


\bibitem{keall2015a}
R.~M. Keall, J.~M. Clayton, and P.~N. Butow.
\newblock Therapeutic {{Life Review}} in {{Palliative Care}}: {{A Systematic
  Review}} of {{Quantitative Evaluations}}.
\newblock {\em Journal of Pain and Symptom Management}, 49(4):747--761, Apr.
  2015. doi: {{%
10\hspace{.1pt}\discretionary{.}{%
}{.}\hspace{.4pt}1016\discretionary{/}{%
}{/}j\hspace{.1pt}\discretionary{.}{%
}{.}\hspace{.4pt}jpainsymman\hspace{.1pt}\discretionary{.}{%
}{.}\hspace{.4pt}2014\hspace{.1pt}\discretionary{.}{%
}{.}\hspace{.4pt}08\hspace{.1pt}\discretionary{.}{%
}{.}\hspace{.4pt}015}}


\bibitem{kerdvibulvech2019}
C.~Kerdvibulvech and S.-U. Guan.
\newblock Affective {{Computing}} for {{Enhancing Affective Touch-Based
  Communication Through Extended Reality}}.
\newblock In S.~Misra, O.~Gervasi, B.~Murgante, E.~Stankova, V.~Korkhov,
  C.~Torre, A.~M.~A. Rocha, D.~Taniar, B.~O. Apduhan, and E.~Tarantino, eds.,
  {\em Computational {{Science}} and {{Its Applications}} \textendash{}
  {{ICCSA}} 2019}, Lecture {{Notes}} in {{Computer Science}}, pp. 351--360.
  {Springer International Publishing}, {Cham}, 2019. doi: {{%
10\hspace{.1pt}\discretionary{.}{%
}{.}\hspace{.4pt}1007\discretionary{/}{%
}{/}978\discretionary{%
}{-}{-}3\discretionary{%
}{-}{-}030\discretionary{%
}{-}{-}24296\discretionary{%
}{-}{-}1\_29}}


\bibitem{lawson2021a}
B.~D. Lawson and K.~M. Stanney.
\newblock Editorial: {{Cybersickness}} in {{Virtual Reality}} and {{Augmented
  Reality}}.
\newblock {\em Frontiers in Virtual Reality}, 2:759682, Oct. 2021. doi: {{%
10\hspace{.1pt}\discretionary{.}{%
}{.}\hspace{.4pt}3389\discretionary{/}{%
}{/}frvir\hspace{.1pt}\discretionary{.}{%
}{.}\hspace{.4pt}2021\hspace{.1pt}\discretionary{.}{%
}{.}\hspace{.4pt}759682}}


\bibitem{mo2022a}
J.~Mo, V.~Vickerstaff, O.~Minton, S.~Tavabie, M.~Taubert, P.~Stone, and
  N.~White.
\newblock How effective is virtual reality technology in palliative care? {{A}}
  systematic review and meta-analysis.
\newblock {\em Palliative Medicine}, p. 026921632210995, May 2022. doi: {{%
10\hspace{.1pt}\discretionary{.}{%
}{.}\hspace{.4pt}1177\discretionary{/}{%
}{/}02692163221099584}}


\bibitem{mok2012a}
E.~Mok, K.-p. Lau, T.~Lai, and S.~Ching.
\newblock The {{Meaning}} of {{Life Intervention}} for {{Patients With
  Advanced-Stage Cancer}}: {{Development}} and {{Pilot Study}}.
\newblock {\em Oncology Nursing Forum}, 39(6):E480--E488, Nov. 2012. doi: {{%
10\hspace{.1pt}\discretionary{.}{%
}{.}\hspace{.4pt}1188\discretionary{/}{%
}{/}12\hspace{.1pt}\discretionary{.}{%
}{.}\hspace{.4pt}ONF\hspace{.1pt}\discretionary{.}{%
}{.}\hspace{.4pt}E480\discretionary{%
}{-}{-}E488}}


\bibitem{nakanishi2014}
H.~Nakanishi, K.~Tanaka, and Y.~Wada.
\newblock Remote handshaking: Touch enhances video-mediated social
  telepresence.
\newblock In {\em Proceedings of the {{SIGCHI Conference}} on {{Human Factors}}
  in {{Computing Systems}}}, pp. 2143--2152. {ACM}, {Toronto Ontario Canada},
  Apr. 2014. doi: {{%
10\hspace{.1pt}\discretionary{.}{%
}{.}\hspace{.4pt}1145\discretionary{/}{%
}{/}2556288\hspace{.1pt}\discretionary{.}{%
}{.}\hspace{.4pt}2557169}}


\bibitem{newell1999}
{Newell}, {Sanson-Fisher}, {Girgis}, and {Ackland}.
\newblock The physical and psycho-social experiences of patients attending an
  outpatient medical oncology department: A cross-sectional study.
\newblock {\em European Journal of Cancer Care}, 8(2):73--82, June 1999. doi:
  {{%
10\hspace{.1pt}\discretionary{.}{%
}{.}\hspace{.4pt}1046\discretionary{/}{%
}{/}j\hspace{.1pt}\discretionary{.}{%
}{.}\hspace{.4pt}1365\discretionary{%
}{-}{-}2354\hspace{.1pt}\discretionary{.}{%
}{.}\hspace{.4pt}1999\hspace{.1pt}\discretionary{.}{%
}{.}\hspace{.4pt}00125\hspace{.1pt}\discretionary{.}{%
}{.}\hspace{.4pt}x}}


\bibitem{niki2019}
K.~Niki, Y.~Okamoto, I.~Maeda, I.~Mori, R.~Ishii, Y.~Matsuda, T.~Takagi, and
  E.~Uejima.
\newblock A {{Novel Palliative Care Approach Using Virtual Reality}} for
  {{Improving Various Symptoms}} of {{Terminal Cancer Patients}}: {{A
  Preliminary Prospective}}, {{Multicenter Study}}.
\newblock {\em Journal of Palliative Medicine}, 22(6):702--707, June 2019. doi:
  {{%
10\hspace{.1pt}\discretionary{.}{%
}{.}\hspace{.4pt}1089\discretionary{/}{%
}{/}jpm\hspace{.1pt}\discretionary{.}{%
}{.}\hspace{.4pt}2018\hspace{.1pt}\discretionary{.}{%
}{.}\hspace{.4pt}0527}}


\bibitem{orts-escolano2016}
S.~{Orts-Escolano}, C.~Rhemann, S.~Fanello, W.~Chang, A.~Kowdle, Y.~Degtyarev,
  D.~Kim, P.~L. Davidson, S.~Khamis, M.~Dou, V.~Tankovich, C.~Loop, Q.~Cai,
  P.~A. Chou, S.~Mennicken, J.~Valentin, V.~Pradeep, S.~Wang, S.~B. Kang,
  P.~Kohli, Y.~Lutchyn, C.~Keskin, and S.~Izadi.
\newblock Holoportation: {{Virtual 3D Teleportation}} in {{Real-time}}.
\newblock In {\em Proceedings of the 29th {{Annual Symposium}} on {{User
  Interface Software}} and {{Technology}}}, pp. 741--754. {ACM}, {Tokyo Japan},
  Oct. 2016. doi: {{%
10\hspace{.1pt}\discretionary{.}{%
}{.}\hspace{.4pt}1145\discretionary{/}{%
}{/}2984511\hspace{.1pt}\discretionary{.}{%
}{.}\hspace{.4pt}2984517}}


\bibitem{pang2005}
S.~M. Pang, K.-S. Chan, B.~P. Chung, K.-S. Lau, E.~M. Leung, A.~W. Leung, H.~Y.
  Chan, and T.~M. Chan.
\newblock Assessing {{Quality}} of {{Life}} of {{Patients}} with {{Advanced
  Chronic Obstructive Pulmonary Disease}} in the {{End}} of {{Life}}.
\newblock {\em Journal of Palliative Care}, 21(3):180--187, Sept. 2005. doi:
  {{%
10\hspace{.1pt}\discretionary{.}{%
}{.}\hspace{.4pt}1177\discretionary{/}{%
}{/}082585970502100311}}


\bibitem{perna2021}
L.~Perna, S.~Lund, N.~White, and O.~Minton.
\newblock The {{Potential}} of {{Personalized Virtual Reality}} in {{Palliative
  Care}}: {{A Feasibility Trial}}.
\newblock {\em American Journal of Hospice and Palliative
  Medicine\textregistered}, 38(12):1488--1494, Dec. 2021. doi: {{%
10\hspace{.1pt}\discretionary{.}{%
}{.}\hspace{.4pt}1177\discretionary{/}{%
}{/}1049909121994299}}


\bibitem{sallnas2004}
E.-L. Sallnäs.
\newblock {\em The effect of modality on social presence, presence and
  performance in collaborative virtual environments}.
\newblock PhD thesis, KTH, 2004.

\bibitem{slater1998}
M.~Slater, A.~Steed, J.~McCarthy, and F.~Maringelli.
\newblock The influence of body movement on subjective presence in virtual
  environments.
\newblock {\em Human factors}, 40:469--77, 10 1998. doi: {{%
10\hspace{.1pt}\discretionary{.}{%
}{.}\hspace{.4pt}1518\discretionary{/}{%
}{/}001872098779591368}}


\bibitem{slater1995}
M.~Slater, M.~Usoh, and Y.~Chrysanthou.
\newblock The influence of dynamic shadows on presence in immersive virtual
  environments.
\newblock {\em VE'95: Selected papers of the Eurographics workshops on Virtual
  environments'951}, 06 1995. doi: {{%
10\hspace{.1pt}\discretionary{.}{%
}{.}\hspace{.4pt}1007\discretionary{/}{%
}{/}978\discretionary{%
}{-}{-}3\discretionary{%
}{-}{-}7091\discretionary{%
}{-}{-}9433\discretionary{%
}{-}{-}1\_2}}


\bibitem{tominari2021}
M.~Tominari, R.~Uozumi, C.~Becker, and A.~Kinoshita.
\newblock Reminiscence therapy using virtual reality technology affects
  cognitive function and subjective well-being in older adults with dementia.
\newblock {\em Cogent Psychology}, 8(1):1968991, Dec. 2021. doi: {{%
10\hspace{.1pt}\discretionary{.}{%
}{.}\hspace{.4pt}1080\discretionary{/}{%
}{/}23311908\hspace{.1pt}\discretionary{.}{%
}{.}\hspace{.4pt}2021\hspace{.1pt}\discretionary{.}{%
}{.}\hspace{.4pt}1968991}}


\bibitem{wang2020}
C.~Y. Wang, M.~Sakashita, U.~Ehsan, J.~Li, and A.~S. Won.
\newblock Again, {{Together}}: {{Socially Reliving Virtual Reality Experiences
  When Separated}}.
\newblock In {\em Proceedings of the 2020 {{CHI Conference}} on {{Human
  Factors}} in {{Computing Systems}}}, pp. 1--12. {ACM}, {Honolulu HI USA},
  Apr. 2020. doi: {{%
10\hspace{.1pt}\discretionary{.}{%
}{.}\hspace{.4pt}1145\discretionary{/}{%
}{/}3313831\hspace{.1pt}\discretionary{.}{%
}{.}\hspace{.4pt}3376642}}


\bibitem{wang2018}
T.~Wang, A.~Molassiotis, B.~P.~M. Chung, and J.-Y. Tan.
\newblock Unmet care needs of advanced cancer patients and their informal
  caregivers: A systematic review.
\newblock {\em BMC Palliative Care}, 17(1):96, Dec. 2018. doi: {{%
10\hspace{.1pt}\discretionary{.}{%
}{.}\hspace{.4pt}1186\discretionary{/}{%
}{/}s12904\discretionary{%
}{-}{-}018\discretionary{%
}{-}{-}0346\discretionary{%
}{-}{-}9}}


\bibitem{warth2019}
M.~Warth, J.~Kessler, F.~Koehler, C.~{Aguilar-Raab}, H.~J. Bardenheuer, and
  B.~Ditzen.
\newblock Brief psychosocial interventions improve quality of life of patients
  receiving palliative care: {{A}} systematic review and meta-analysis.
\newblock {\em Palliative Medicine}, 33(3):332--345, Mar. 2019. doi: {{%
10\hspace{.1pt}\discretionary{.}{%
}{.}\hspace{.4pt}1177\discretionary{/}{%
}{/}0269216318818011}}


\end{thebibliography}
\end{document}